\documentclass[sigconf,nonacm]{acmart}
\AtBeginDocument{%
  }

\setcopyright{acmlicensed}
\copyrightyear{2025}
\acmYear{2025}
\acmConference[C+J '25]{Computation + Journalism Symposium 2025}{December 11--12,
  2025}{Miami, FL}

\usepackage{listings}
\usepackage{xcolor}
\begin{document}

\title{LLM-Assisted News Discovery in High-Volume Information Streams: A Case Study}

\author{Nick Hagar}
\authornote{Corresponding author: nicholas.hagar@northwestern.edu}

\author{Ethan Silver}

\author{Clare Spencer}

\author{Nicholas Diakopoulos}

\affiliation{%
  \institution{Northwestern University}
  \city{Evanston}
  \state{IL}
  \country{USA}
}

\renewcommand{\shortauthors}{Hagar et al.}

\begin{abstract}
  Journalists face mounting challenges in monitoring ever-expanding digital information streams to identify newsworthy content. While traditional automation tools gather information at scale, they struggle with the editorial judgment needed to assess newsworthiness. This paper investigates whether large language models (LLMs) can serve as effective first-pass filters for journalistic monitoring. We develop a prompt-based approach encoding journalistic news values---timeliness, impact, controversy, and generalizability---into LLM instructions to extract and evaluate potential story leads. We validate our approach across multiple  models against expert-annotated ground truth, then deploy a real-world monitoring pipeline that processes trade press articles daily. Our evaluation reveals strong performance in extracting relevant leads from source material ($F1=0.94$) and in coarse newsworthiness assessment ($\pm$1 accuracy up to 92\%), but it consistently struggles with nuanced editorial judgments requiring beat expertise. The system proves most valuable as a hybrid tool combining automated monitoring with human review, successfully surfacing novel, high-value leads while filtering obvious noise. We conclude with practical recommendations for integrating LLM-powered monitoring into newsroom workflows that preserves editorial judgment while extending journalistic capacity.
\end{abstract}




\maketitle

\section{Introduction}
Information overload is an acute issue for beat reporters who must continuously monitor specialized streams of information---whether tracking research developments in science journalism \cite{nishal_understanding_2024}, processing audience contributions in engagement journalism \cite{wang_journalistic_2021}, or covering local events with severely limited resources \cite{mcchesney_farewell_2012}. The fundamental challenge is not merely accessing information, but efficiently filtering it according to complex, often implicit journalistic criteria to identify what merits further investigation and coverage.

News organizations have long turned to automation to help manage these information streams \cite{marconi_future_2017,liu_reuters_2016,diakopoulos_automating_2019}. However, traditional automation approaches---typically based on keyword matching, pattern detection, or rule-based heuristics---excel at scale and speed but struggle with the nuanced editorial judgment required to assess newsworthiness. They can surface large volumes of potentially relevant content but offer limited help in determining what actually deserves a journalist's attention, often simply shifting the bottleneck from information discovery to information triage \cite{atreja_understanding_2023}.

Whereas some approaches to this problem have deployed standard machine learning to help with filtering and ranking \cite{nishal_understanding_2024, diakopoulos_towards_2021}, recent advances in large language models (LLMs) suggest a new and potentially effective approach. LLMs demonstrate remarkable capability in understanding context, following complex instructions, and making nuanced textual assessments \cite{brown_language_2020}. When properly prompted, these models can perform sophisticated text classification and analysis tasks that approach human-level performance \cite{zhang_towards_2023}, and early experiments have shown promise in helping journalists interpret documents \cite{nishal_understanding_2024, petridis2023anglekindling-f41}. This raises an intriguing possibility: Could LLMs serve as effective editorial filters, encoding basic journalistic judgment into the monitoring process itself?

This paper investigates the potential of LLMs to assist journalists in monitoring, filtering, and prioritizing information streams relevant to their beats. We propose and evaluate a prompt-based approach that leverages LLMs to identify newsworthy content according to explicit journalistic criteria. Using the beat of ``generative AI in newsrooms'' as our case study\footnote{https://generative-ai-newsroom.com/}, we develop a systematic approach for prompt engineering based on established news values \cite{diakopoulos_computational_2020}, implement an automated monitoring pipeline using open-source workflow tools, and conduct both controlled evaluations and real-world deployment testing with an active beat reporter.

With this work, we first demonstrate that LLMs can accurately extract newsworthy items from diverse information sources while maintaining reasonable alignment with human editorial judgment. Second, we identify the boundaries of automated curation, showing where human expertise remains essential for contextual assessment and beat-specific knowledge. Third, we provide practical recommendations for integrating LLM-powered monitoring into journalistic workflows. In doing so, we aim to move toward concrete, deployable solutions that can help resource-constrained newsrooms manage information overload without sacrificing editorial standards.

\section{Background}
Reporters often need to scan a wide variety of incoming information sources---from wires and press releases to social feeds---to catch emerging stories. This relentless influx leads to information overload \cite{atreja_understanding_2023}. In response, a core journalistic task is \textit{curation}---deciding where to focus limited time and attention amid a deluge of potential stories \cite{bruns_gatewatching_2003, thorson_curated_2016}. This gatekeeping role appears across many journalistic functions. Science reporters, for example, must choose which research papers are newsworthy enough to highlight, and the simultaneous expansion of scientific publishing and shrinking of journalistic resources makes it challenging for them to find newsworthy leads. In response, they increasingly seek tools to help filter and prioritize which scientific findings merit coverage \cite{nishal_understanding_2024}. In the context of audience engagement, many newsrooms solicit tips, comments, and user-generated content from their readers. Engagement editors must manually decide which among thousands of comments or tips might be worth pursuing as a story, underscoring the need for better filtering support \cite{wang_journalistic_2021}. And at the local level, the extreme resource constraints facing many community-based news publishers limits what kinds of events they can cover and which perspectives they are able to capture in their reporting \cite{mcchesney_farewell_2012,mcchesney_problem_2003}.

Because newsroom resources are limited, journalists have turned to automation and computational tools to assist with monitoring information streams. For financial news, major outlets use algorithms to detect news and even draft basic stories \cite{marconi_future_2017, peiser_rise_2019}. More broadly, reporters often employ automated tools to monitor online resources and alert them of any important updates \cite{ciobanu_klaxon_2017, liu_reuters_2016,robinson_keep_2019}. Some organizations subscribe to commercial alert services (e.g., Dataminr) that alert journalists when keywords or unusual activity patterns suggest something newsworthy might be happening \cite{diakopoulos_automating_2019}. These tools are helpful for catching fast-moving events as they first surface online. In each of these cases, automation now plays a routine role in newsgathering \cite{diakopoulos_automating_2019,atreja_understanding_2023}. The goals of automated tools are often to extend the journalists’ reach (by monitoring data sources at scale) and to accelerate the news cycle (by producing initial drafts or alerts that reporters can then refine). When effectively integrated, such tools can help newsrooms overcome resource constraints and not miss important stories.

However, such tools also run the risk of reproducing or amplifying the very information overload that they are designed to help manage. Online information environments still require a lot of manual triage from journalists \cite{atreja_understanding_2023}; a high volume of automatically surfaced items (tweets, alerts, documents, etc.) still requires human judgment to separate signal from noise.

Recent research demonstrates that state-of-the-art language models have become quite proficient at understanding and classifying text when guided with the right prompts. These models achieve impressive performance on a range of NLP tasks without task-specific training \cite{brown_language_2020}. For example, with only a few examples provided in a prompt, GPT-3.5 can equal or even outperform supervised models on certain text classification and fact-verification benchmarks \cite{zhang_towards_2023}. Applying generative AI in this way opens the door to LLM-powered monitoring systems that go beyond raw surveillance and add a layer of editorial judgment. The model can not only pull out data, but also summarize and prioritize it. Early experiments are encouraging: LLMs can follow explicit instructions to perform complex tasks like summarization, sentiment analysis, and fact-checking of text \cite{zhang_towards_2023}. In journalism, prototype systems have used GPT-4 to help reporters digest technical documents \cite{nishal_-jargonizing_2024}.

In short, LLMs may offer a way to encode a first layer of journalistic judgment into the monitoring process itself. They bring the ability to interpret text more like a human reader, simulating an understanding of meaning, implications, and news values. With a well-crafted prompt, an LLM can be instructed to prioritize certain topics, tones, or criteria. This represents a qualitative leap from simple automation toward a form of AI-assisted curation. This work explores such an ``LLM-as-curator'' approach for monitoring, summarizing, and prioritizing timely trade press articles. In doing so, we propose an approach to formalize and evaluate AI-assisted curation in the newsroom, paving the way for practical uptake in reporting workflows and further exploration of model capabilities. 

\section{Proposed Approach}
We propose a prompt-based approach that leverages widely available large language models to assist journalists in monitoring, filtering, and prioritizing information streams relevant to their beats. The system monitors relevant publications, extracts newsworthy details, and presents only the most promising leads for further investigation. To test this approach, we focus on the beat of \textit{generative AI in the newsroom}---a timely and rapidly evolving topic where the authors possess domain expertise, enabling rigorous evaluation of the system's outputs.

Our methodology proceeds in two phases: prompt engineering to capture journalistic news values in LLM instructions, followed by pipeline development to test the approach in a real-world monitoring scenario.

\subsection{Prompt Engineering}
The first phase focused on developing and validating a prompt that could effectively guide LLMs to identify newsworthy content according to journalistic criteria. We began by assembling a ground truth dataset of trade press articles about AI in journalism, collected over a three-month period from prominent industry sources including the Google News Initiative, the Online News Association, and the Reuters Institute \cite{silver_keeping_2025}. This dataset contained a corpus of 63 articles reflecting 144 use cases.\footnote{https://airtable.com/appjOwPCsQXygqcIs/shriLjLNH2uqT788L/tblQ4IMeaKdyMqvKm
}

Through iterative analysis of the ground truth data and relevant use cases, and drawing from prior work on news discovery and newsworthiness \cite{diakopoulos_towards_2021,diakopoulos_computational_2020}, we identified four facets that our prompt should capture: timeliness, impact, controversy, and general audience appeal.

We developed our prompt through an iterative refinement process, beginning with initial templates based on established journalistic news values and refining based on performance against our ground truth data. Each iteration involved testing the prompt on a subset of articles, analyzing failures and edge cases, and adjusting the instructions to better align with human judgments. The final prompt structure included explicit instructions for each facet and specific output formatting requirements to ensure consistent, actionable summaries. The final prompt appears in Appendix A. 

We evaluated the refined prompt using multiple LLMs from OpenAI: GPT-4o (2024-08-06 snapshot), GPT-4.1 (2025-04-14 snapshot), GPT-5 (2025-08-07 snapshot), o4-mini (2025-04-16 snapshot), and o3 (2025-04-16 snapshot). All models were tested with default parameters. For this evaluation, three of the authors annotated a randomly-selected sample of 20 articles from our corpus, from which they extracted 75 specific \textit{use cases} to establish baseline relevance judgments.. We compared model outputs---both identified use cases and newsworthiness ratings---against these human annotations. We measured performance using standard metrics (MAE, $\pm1$ accuracy, precision, recall, and F1 score). To establish a human baseline, we calculated inter-annotator agreement among the authors using Cohen's kappa.

\subsection{Pipeline Development}
Because it offers an open source version, we selected n8n as our workflow automation platform\footnote{\url{https://github.com/n8n-io/n8n}}. The pipeline architecture consists of four main components. First, RSS feed monitors track Google Alerts using keyword filters to identify potentially relevant content about AI and journalism (see the full list of keywords in Appendix B). Second, a parsing module extracts article URLs from each RSS feed, then collects the text of each article's web page. Third, the LLM processing module sends each article's text to the OpenAI API along with our prompt, receiving structured assessments in return. Finally, an output management module formats the results and exports them to an Airtable spreadsheet. The n8n workflow for this pipeline is available in JSON format on the project repository.\footnote{\url{https://github.com/NHagar/llm_use_cases_monitoring}}

We configured the pipeline to run once per day at a consistent time, balancing timeliness with API costs. Each processed article generates a structured output following the schema shown in Listing \ref{lst:schema}---an overall summary of the article, a usefulness rating for the article (1-5 scale), and a list of identified use cases. Each use case item contains a name, description, key details, and a newsworthiness rating (1-5 scale).

\begin{lstlisting}[float=h, caption={Schema for the outputs of LLM article processing}, label={lst:schema}]

UseCase:
    name: str
    description: str
    ai_model_used: Optional[str]
    strengths: str
    challenges: str
    newsroom_impact: str
    link_to_demo: Optional[str]
    is_original: bool
    comparison_to_other_use_cases: Optional[str]
    newsworthiness_rating: float

Article:
    summary: str
    usefulness_rating: float
    use_cases: list[UseCase]
\end{lstlisting}

To evaluate the pipeline's real-world effectiveness, we conducted a one-week field test in which the system monitored live information streams while one of the authors---an active reporter on this beat---provided quantitative and qualitative feedback. This evaluation focused on three key questions: How newsworthy are the surfaced use cases, on a scale of 1--5? What additional details would be required to report on these use cases? Are these use cases original, based on the author's prior knowledge?

\section{Results}
\subsection{Prompt Evaluation}
\paragraph{Use-case identification.}
We first evaluate how well each model, when guided by the same prompt, identifies the ground-truth use cases extracted during labeling (75 total). Table~\ref{tab:coverage} reports true positives (TP), false positives (FP), false negatives (FN), and derived precision/recall/F1. The models exhibit distinct operating characteristics: \texttt{o3} attains the strongest balanced performance (F1{=}0.94; Precision{=}0.91, Recall{=}0.96), missing only 3 of 75 use cases while keeping false alarms modest. \texttt{o4-mini} yields the highest precision (0.97) with fewer false positives (FP{=}2) at the expense of recall (0.87; FN{=}10). \texttt{gpt-4o} trades off recall most heavily (0.68), and \texttt{gpt-5} produces the most false positives (FP{=}24; Precision{=}0.71).

\begin{table}[ht]
\centering
\small
\begin{tabular}{lrrrrrrrr}
\hline
Model & TP & FP & FN & FP\% & FN\% & Precision & Recall & F1 \\
\hline
\texttt{o3}       & 72 &  7 &  3 &  8.9 &  4.0 & 0.911 & 0.960 & 0.935 \\
\texttt{o4-mini}  & 65 &  2 & 10 &  3.0 & 13.3 & 0.970 & 0.867 & 0.915 \\
\texttt{gpt-4o}   & 51 &  7 & 24 & 12.1 & 32.0 & 0.879 & 0.680 & 0.767 \\
\texttt{gpt-4.1}  & 65 & 11 & 10 & 14.5 & 13.3 & 0.855 & 0.867 & 0.861 \\
\texttt{gpt-5}    & 60 & 24 & 15 & 28.6 & 20.0 & 0.714 & 0.800 & 0.755 \\
\hline
\end{tabular}
\caption{Use-case identification coverage relative to 75 ground-truth use cases.}
\label{tab:coverage}
\end{table}

\paragraph{Newsworthiness rating agreement.}
We next compare model newsworthiness scores (1--5) against the mean of three human annotators for the 75 labeled use cases.\footnote{Per-model sample sizes ($N$) reflect only items for which the model output a usable rating.} Human ratings show notable subjectivity (inter-annotator Cohen's $\kappa$ between 0.034 and 0.290), which constrains the attainable agreement. Against the human mean, models exhibit moderate linear association (Pearson $r \approx 0.23$--$0.52$) but low exact-match accuracy (from 0.143 to 0.462) and consistently higher average scores than humans. Still, $\pm 1$-point accuracy is high for most models (from 0.716 to 0.923), indicating utility for coarse-grained prioritization even when absolute scores diverge.

\begin{figure}
    \centering
    \includegraphics[width=0.95\linewidth]{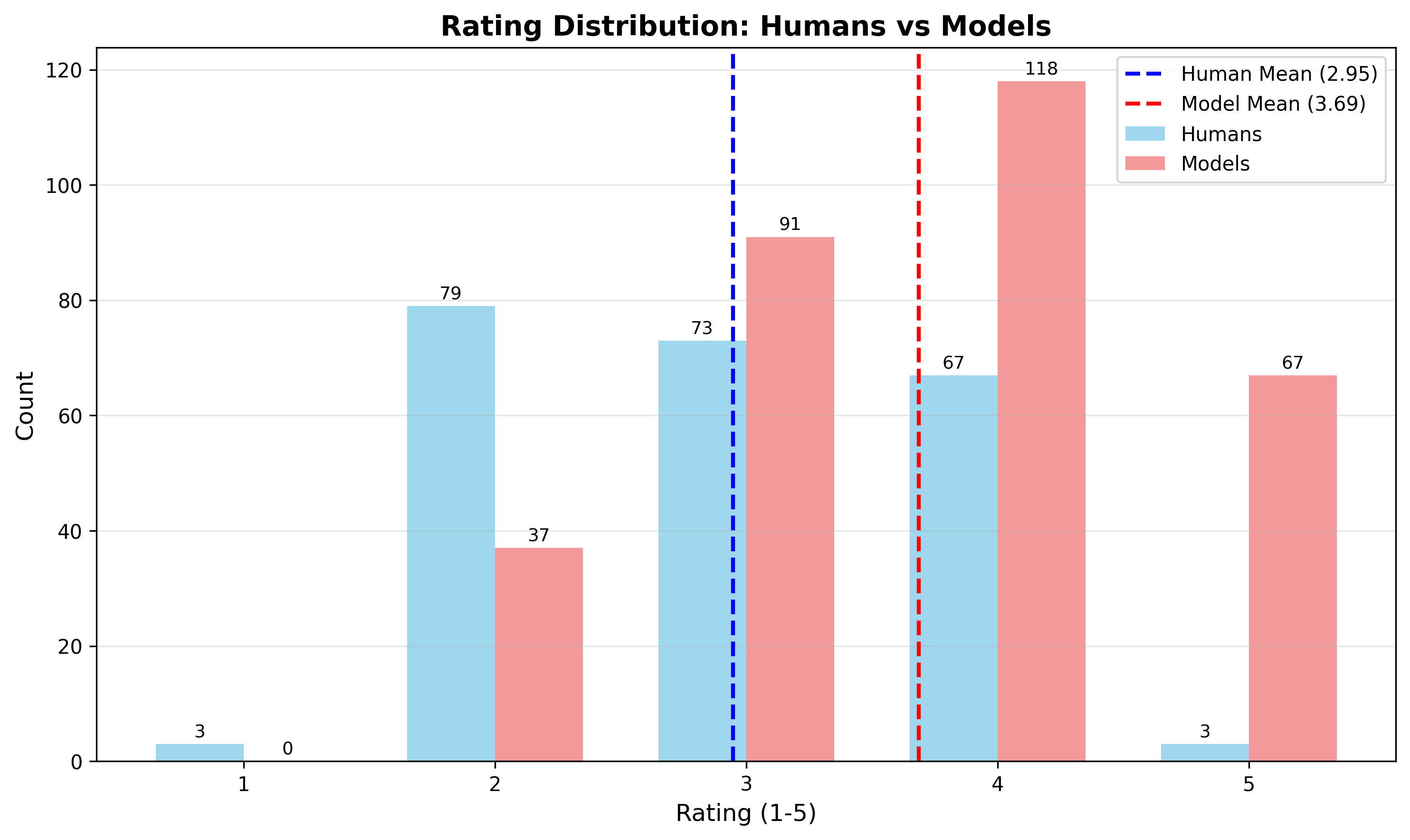}
    \caption{Rating distributions (1--5). LLMs tend to rate higher on average than human annotators.}
    \label{fig:means}
\end{figure}

Figure~\ref{fig:means} summarizes rating distributions; Table~\ref{tab:reg} reports evaluation metrics computed against the average human rating.

\begin{table}[t]
\centering
\small
\begin{tabular}{lrrrrrr}
\hline
Model & $N$ & MAE & RMSE & $R^2$ & Pearson $r$ & Accuracy / $\pm$1 Acc \\
\hline
\texttt{o3}      & 72 & 0.792 & 1.005 & $-1.244$ & 0.516 & 0.444 / 0.861 \\
\texttt{o4-mini} & 65 & 0.713 & 0.903 & $-0.810$ & 0.459 & 0.462 / 0.923 \\
\texttt{gpt-4.1} & 67 & 1.234 & 1.404 & $-3.609$ & 0.274 & 0.164 / 0.716 \\
\texttt{gpt-4o}  & 49 & 1.177 & 1.336 & $-3.152$ & 0.234 & 0.143 / 0.755 \\
\texttt{gpt-5}   & 60 & 0.922 & 1.143 & $-1.805$ & 0.481 & 0.367 / 0.800 \\
\hline
\end{tabular}
\caption{Agreement with the \emph{average} human rating. Accuracy: exact score match on the 1--5 scale; $\pm$1 Acc: within one point. Negative $R^2$ indicates poor calibration to the human scale despite moderate correlation.}
\label{tab:reg}
\end{table}

\paragraph{Summary.}
For this prompt, o3 provides the best balance of coverage and precision for identifying use cases, while o4-mini offers the most conservative filter with minimal false positives. For rating newsworthiness, all models show moderate correlation with humans but low exact agreement and a tendency to inflate the score, consistent with the low human--human agreement that characterizes this subjective task.

\subsection{Pipeline Deployment}
Based on the results of the prompt evaluation, we deployed our pipeline using o3 for language model processing. Over the course of one week, the pipeline processed 89 articles surfaced by Google Alerts, extracting 274 use cases. One of the authors, an active reporter covering this beat, annotated a random sample of 20 use cases extracted into a separate spreadsheet, representing 7\% of the pipeline's total output.

\paragraph{Newsworthiness rating agreement.}
Average newsworthiness scores were closely aligned (pipeline $M{=}3.27$ vs.\ human $M{=}3.47$), but agreement at the record level was limited (exact{=}26.7\%, $\pm$1{=}46.7\%).

\paragraph{Pipeline errors.}
At the record level, the pipeline demonstrated several consistent weaknesses across the annotated sample. First, because the prompt operates at the article level, it does not capture cases where multiple articles cover the same use case. This led to 4 duplicate use cases in our sample. Second, in cases where the LLM rated use cases as \textit{more} newsworthy than human annotation, the disagreement consistently came from a lack of specificity---the LLM latched onto vague use case descriptions that could prove interesting with further reporting (e.g., ``AI-driven investigative tools''), whereas the author would require far more concrete details to consider pursuing a lead. Finally, in cases where the LLM rated use cases as \textit{less} newsworthy than human annotation, the LLM appeared to under-value real-world impact, relative to technical innovation.

\paragraph{Triage value.}
Using a high-newsworthiness threshold ($\geq 4$), the pipeline achieved balanced triage performance (precision{=}0.50, recall{=}0.50, $F_{1}{=}0.50$). Notably, 3 use cases (15.8\%) were jointly rated ``high'' by both the model and the author and were \emph{novel} to the author. These shared two traits: explicit newsroom impact (workflows, adoption paths) and specific operational details (tools, deployments, policies). These patterns suggest that, even with modest average-level agreement, the system does lift a minority of high-value, novel candidates into an editor’s attention stream.

\paragraph{Follow-up information required.}
The author's qualitative feedback indicated that while the LLM populated descriptive fields like strengths, challenges, and newsroom impact, the generated text often lacked the specificity required for reporting. This gap between the automated summary and journalistic need is reflected in the five common areas where the author required additional information to pursue a lead: (i) concrete case studies, (ii) sustainability and business model details, (iii) technical specifics, (iv) deployment scope and timing, and (v) named sources/points of contact.

\paragraph{Throughput \& cost.}
A typical run completed in about 12 minutes with low inference cost ($\approx$\$0.15 per run), indicating that LLM-powered monitoring is practical at newsroom scales.

\section{Discussion}
In this study, we demonstrate that LLMs, when carefully prompted, show potential as first-pass filters for journalists who need to stay apprised of constantly-updating streams of information. We first show that frontier models (in particular, OpenAI's o3 and o4-mini) can accurately extract individual items of interest from published material across news articles, technical reports, and press releases. In addition, these models demonstrate coarse alignment with human newsworthiness ratings of extracted items, as shown in their high $\pm1$ accuracy. 

At the same time, our findings emphasize the continued need for human expertise in evaluating newsworthiness. As our pilot deployment made clear, the benefit of LLMs comes primarily from monitoring incoming information, enforcing a standardized schema, and filtering out the most obvious noise. For more fine-grained judgments---deciding whether a use case was, for example, novel or worth dedicating reporting time to---the reporter in our deployment found less utility from relying on the LLM, reinforcing prior work's emphasis on the necessity of editorial judgment in news discovery \cite{diakopoulos_computational_2020}. In particular, the LLM consistently overvalued technically novel but vaguely described use cases while undervaluing concrete implementations with clear real-world impact but less technical sophistication---a pattern that inverts the priorities of beat reporters who need actionable, specific leads over speculative possibilities.

In broader newsroom use across multiple beats or streams of information, our findings emphasize the value of a \textit{hybrid monitoring approach} \cite{atreja_understanding_2023}. Such a system leverages computational tools like feed monitoring and LLMs for their strengths (scale and speed), while ensuring human control throughout the process. In our approach, this takes the form of \textit{careful prompt engineering} via clearly-defined ground truth data, iterative development, and consensus across stakeholders around success and failure conditions; \textit{transparent inputs and outputs} in the form of an open-source workflow and standard data formats; and \textit{human editorial review}. These practices can help ensure that an LLM-powered monitoring system extends a reporter's ability to focus on high-quality leads, rather than subsuming their judgment or overwhelming them with noise. 

\subsection{Limitations}
Because this system leverages LLMs, the reported outputs are all probabilistic. Repeated queries with the same prompts may return different use cases, ratings, or justifications. In addition, our pipeline deployment uses Google Alerts RSS feeds as its source of candidate articles. While these feeds help us monitor a broad slice of material published on the web, it is possible that a more targeted approach to curating candidate articles would change the quality of the information available to the LLM. Finally, while we propose general principles that can be applied to multiple beats, the case study tested here was intentionally narrow. Other monitoring scenarios may have specific data or prompting requirements that are not captured in this work. 

\subsection{Conclusion}
This work presents a practical approach for integrating large language models into journalistic monitoring workflows, demonstrating that carefully prompted LLMs can serve as effective first-pass filters for information streams. Through our case study of monitoring generative AI use cases in newsrooms, we show that frontier models can accurately extract newsworthy items from diverse sources while maintaining coarse alignment with human editorial judgment. More importantly, our deployment reveals the boundaries of automated curation: while LLMs excel at scale and standardization, they cannot replace the contextual knowledge, beat expertise, and nuanced judgment that journalists bring to newsworthiness assessment. This finding reinforces that the path forward is not full automation but rather augmentation---systems that amplify journalistic capacity without supplanting journalistic values.

\begin{acks}
This work was conducted with support from the Knight Foundation and the Early Research Experience Awards program at Northwestern University's School of Communication.
\end{acks}

\bibliographystyle{ACM-Reference-Format}
\bibliography{sample-base}

\appendix
\section{Monitoring Prompt}
You are an expert researcher at the intersection of journalism and artificial intelligence. When provided with a resource in the next message (e.g., article, website, tool, or paper), first read the full resource thoroughly, then generate a structured report in clear, professional language. The output should be concise, insightful, and formatted to serve journalists, editors, academics, and product teams. Use bullet points where appropriate, and structure the response clearly.  
  
---  
  
\#\#\# 1. Summary  \\
Briefly summarize the resource. Clearly state:  \\
- What the resource/project is and who created it  \\
- Its main purpose and key insights  \\
- Why this resource is relevant or significant right now  
  
\#\#\# 2. Usefulness Rating  \\
A good resource will be easily-understood by a professional journalist and contain explicit references to journalism/newsmaking. The content of the resource must be directly applicable to the journalism profession and the newsroom. Critically rate its usefulness for journalists on a scale from 1 to 10. 1 is not relevant to journalists at all, while 10 means absolutely essential. Include in your score the credibility of the source. Provide 1 sentence justifying your score.

\#\#\# 3. Based on the information provided in the resource, identify and list the major specific, use cases for **generative** artificial intelligence (AI) in the newsroom mentioned in the article. The use cases must be explicitly mentioned in the resource.  
  
For **each** use case listed above, answer the following in bullet form (answer all, or indicate ``not specified''):  

\begin{itemize}
    \item What AI model(s) or techniques are used?
    \item What aspects or features worked particularly well?
    \item What challenges or limitations were noted?
    \item What was its impact on the newsroom and its content?
    \item Is there a public demo, dataset, or project available? Provide a link if applicable.
    \item Does it present original ideas, credible evidence, or practical applications?
    \item How does it compare to similar use cases, tools, or approaches?
\end{itemize}
  
Then, evaluate if each use case identified is newsworthy based on the following criteria. A resource has generally high newsworthiness if:  
  
\begin{itemize}
    \item It is relevant to contemporary issues in society.
    \item It has the potential to impact many people, positively or negatively.
    \item It has the potential for controversy or debate.
    \item It can be understood by a general audience, and this criterion should be weighted twice as heavily as the others.
\end{itemize}

The following questions for general useworthiness should be used as a guide, not a checklist. Be critical in your judgment.  
  
\begin{itemize}
    \item Is it true?
    \item Is it new?
    \item Is it interesting?
    \item Is it important?
\end{itemize}

Consider all these criteria separately. Remember, the use case must be specifically newsworthy in regards to generative AI in the newsroom. Then, rate the overall newsworthiness of the use case on a scale from 1 (not newsworthy) to 5 (highly newsworthy).  
  
---  
  
**Tone:** Be concise and analytical, but accessible. Use a clear and professional tone appropriate for journalists, editors, and academics.

\section{Google Alert Keywords}
Below is the full list of keyword combinations used to create RSS feeds for monitoring relevant articles.

\begin{itemize}
    \item AI, Journalism
    \item AI, Journalism, Use Cases
    \item AI, Newsroom, Use Cases
    \item Generative AI, Journalism
    \item Generative AI, Journalism, Use Cases
    \item Generative AI, Newsroom
    \item Generative AI, Newsroom, Use Cases
\end{itemize}

\end{document}